# EFFECTS OF BEAM SIZE AND PULSE DURATION ON THE LASER DRILLING PROCESS

**Nazia Afrin, Pengfei Ji, J. K. Chen and Yuwen Zhang[1]**
Department of Mechanical and Aerospace Engineering
University of Missouri
Columbia, MO 65211, USA

**ABSTRACT**

A two-dimensional axisymmetric transient laser drilling model is used to analyze the effects of laser beam diameter and laser pulse duration on the laser drilling process. The model includes conduction and convection heat transfer, melting, solidification and vaporization, as well as material removal resulting from the vaporization and melt ejection. The validated model is applied to study the effects of laser beam size and pulse duration on the geometry of the drilled hole. It is found that the ablation effect decrease with the increasing beam diameter due to the effect of increased vaporization rate, and deeper hole is observed for the larger pulse width due to the higher thermal ablation efficiency.

## 1. INTRODUCTION

Laser drilling (LD) can find its applications in automotive, aerospace, and material processing [1-4]. The laser material interaction and its applications have undergone much study in recent years. In the laser drilling problem, a laser beam is produced and delivered to the target material that absorbs a fraction of the incident laser energy. This energy is then conducted into the target material and heating occurs, resulting in melting and vaporization of the target material. A time- and position-dependent vapor pressure exerts on the melt surface which results in a time- and position-dependent saturation temperature at the liquid surface. The resultant recoil surface pressure pushes the liquid out of the developing cavity and material is removed by a combination of vaporization and melt expulsion. Laser drilling process includes heat transfer into the metal, thermodynamics of phase change and incompressible fluid flow due to the imposed pressure with a free boundary at the melt/vapor interface, and a moving boundary at the melt/solid interface due to the presence of melting process. The moving melt-solid interface and moving liquid-vapor interface result in a special type of problem called Stefan problem with two moving boundaries, where Stefan boundary conditions are enforced.

Laser drilling process requires clear understanding of fundamental physics for better control and increasing the efficiency of the process. Sometime it is difficult to attain small and accurate diameter of the holes on the workpiece. Due to the small size of the hole and melting region, even though the presence of the laser beam itself, it is almost impossible to measure regularly temperature, pressure as well as flow condition above the melt region. Moreover, vaporization, phase change and gas dynamics are important in LD process. Numerical simulation for LD process helps understanding the complex phenomena. Two-dimensional axisymmetric model was proposed by Ganesh et al. [5] to consider resolidification of the molten metal, transient drilled hole development and expulsion of the liquid metal in the LD process.

A number of studies of the laser drilling process can be found in the literature [6-10]. Most of these studies considered one dimensional and were primarily based on thermal arguments. Von Allmen [6] used one-dimensional theoretical model for rate of vaporization and liquid expulsion to calculate the velocity and the efficiency of laser drilling process as a function of absorbed intensity. Chan and Mazumder [7] developed a one dimensional steady-state model which provided close form of analytical solution for damage by liquid vaporization and expulsion. Kar and Mazumder [11] formulated a two-dimensional axisymmetric model that neglected the fluid flow of the target material in melt layer.

The effects of fluid flow and convection were considered on the melted pools in welding [12, 13]. Chan et al. [14] developed a two-dimensional transient model where the solid-liquid interface was considered as a part of the solution and the surface of the melt pool was assumed to be flat to simplify the application of the boundary conditions. A Gaussian temperature distribution as boundary condition was imposed on the top surface, and surface tension and buoyancy driving forces are accounted for in Kou and Wang's study [15]. Two-dimensional axisymmetric transient model of LD problem considering conduction and advection heat transfer in the solid and liquid metal, free flow of liquid melt and its expulsion and the evolution of latent heat of fusion over a temperature range was modeled to track the solid–liquid and liquid-vapor interfaces with different thermophysical properties [17, 18]. Zhang and Faghri [19] developed a thermal model of the melting and vaporization phenomena in the laser drilling process based on energy balance analysis at the solid-liquid and liquid-vapor interfaces. The predicted material removal rate agreed very well with the experimental data. They found that effect of heat loss through conduction was insignificant on the vaporization, and the locations of melting front is significantly affected by conduction heat loss for low laser intensity and longer pulse.

---
[1] Corresponding author. Email: zhangyu@missouri.edu

There are many parameters that have influences on the laser drilling process and thereby the quality that can be achieved. Laser wavelength, laser pulse width and peak power are most influential among of them. The objective of this paper is to study the effects of laser beam diameter and pulse width on the LD process.

## 2. ANALYTIC MODEL

A schematic representation of the processes occurs in LD in shown in Fig. 1. In this model, a laser beam is produced and directed towards a metal target which absorbs a fraction of the light energy that results in melting and vaporization of the metal target. The recoil pressure resulting vaporization pushes the liquid away in the radial direction. In other words, the material is removed by the combination of the vaporization and liquid expulsion. This model includes heat conduction and convection, fluid dynamics of melting flow with free surface at the liquid-vapor interface, and vaporization at the melting surface and resulting melting surface temperature and pressure profiles.

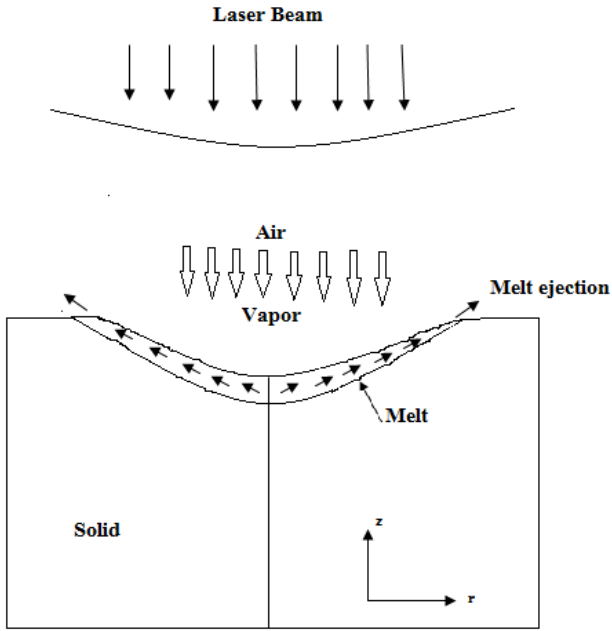

**Figure 1** Schematic diagram of laser drilling process

The hydrodynamical equations are applicable in the liquid regions. The non-dimensional governing equations in 2-D axisymmetric cylindrical coordinate system are

$$\frac{\partial U}{\partial R} + \frac{\partial V}{\partial R} + \frac{U}{R} = 0 \quad (1)$$

$$\frac{\partial U}{\partial T} + U\frac{\partial U}{\partial R} + V\frac{\partial U}{\partial Z} = \Pr[\frac{\partial^2 U}{\partial R^2} + \frac{\partial^2 U}{\partial Z^2} + \frac{1}{R}\frac{\partial U}{\partial R} - \frac{U}{R^2}] - \frac{\partial P}{\partial R} \quad (2)$$

$$\frac{\partial V}{\partial T} + U\frac{\partial V}{\partial R} + V\frac{\partial V}{\partial Z} = \Pr[\frac{\partial^2 V}{\partial R^2} + \frac{\partial^2 V}{\partial Z^2} + \frac{1}{R}\frac{\partial V}{\partial R}] - \frac{\partial P}{\partial R} - G \quad (3)$$

where non-dimensional variables are

$$U = \frac{ud_0}{\alpha_1}, V = \frac{vd_0}{\alpha_1}, G = \frac{gd_0^3}{\alpha_1^2}$$

$$P = \frac{pd_0^2}{\rho\alpha_1^2}, R = \frac{r}{d_0}, Z = \frac{z}{d_0}, \Pr = \frac{\mu}{\rho\alpha_1}, T = \frac{t\alpha_1}{d_0^2} \quad (4)$$

where g, σ and Pr are the gravitational acceleration, surface tension coefficient, and Prandtl number, respectively. Characteristics length (laser beam diameter) and the thermal diffusivity of the liquid melt are represent by $d_0$ and $\alpha_1$, respectively.

The energy equation is solved as an advection-diffusion equation that accounts phase change phenomena via a temperature transforming model. For a single time step, temperature field is obtained for a given fixed velocity. The important treatment of the LD problem is to consider melting surface as a free surface and its location is tracked by the volume of fluid (VOF) method. The volume of fluid function, F is defined as unity for full fluid cell and null (zero) for the empty cell. A donor-acceptor flux approximation method is used to handle the VOF function (F) that cannot be handled by the finite volume method [20]. The governing equation for F is given by

$$\frac{\partial F}{\partial T} + U\frac{\partial F}{\partial R} + V\frac{\partial F}{\partial Z} = 0 \quad (5)$$

The nondimensional energy equation in the cylindrical coordinate system is

$$\frac{\partial(CT')}{\partial T} + \frac{\partial(UCT')}{\partial R} + \frac{\partial(VCT')}{\partial Z}$$
$$= \frac{\partial}{\partial R}(K\frac{\partial T'}{\partial R}) + \frac{\partial}{\partial Z}(K\frac{\partial T'}{\partial Z}) + \frac{1}{R}\frac{\partial}{\partial R}(KT') + B \quad (6)$$

where

$$C(T) = \begin{bmatrix} C_{sl} & (T' < -\delta T') \\ \frac{1}{2}(1+C_{sl}) + \frac{1}{2Ste\delta T'} & (-\delta T' \leq T' \leq \delta T') \\ 1 & (T' > \delta T') \end{bmatrix}$$

$$B = -[\frac{\partial S}{\partial T} + \frac{\partial(US)}{\partial R} + \frac{\partial(VS)}{\partial Z}]$$

$$S(T^{'}) = \begin{bmatrix} C_{sl}\delta T^{'}, & T^{'} < -\delta T^{'} \\ \frac{1}{2}(1+C_{sl})\delta T^{'} + \frac{1}{2Ste}, & -\delta T^{'} \leq T^{'} \leq \delta T^{'} \\ C_{sl}\delta T^{'} + \frac{1}{Ste}, & T^{'} > \delta T^{'} \end{bmatrix}$$

$$K(T^{'}) = \begin{bmatrix} K_{sl}, & T^{'} < -\delta T^{'} \\ K_{sl} + \frac{(1-K_{sl})(T^{'}+\delta T^{'})}{2\delta T^{'}}, & -\delta T^{'} \leq T^{'} \leq \delta T^{'} \\ 1, & T^{'} > \delta T^{'} \end{bmatrix} \quad (7)$$

The nondimensional variables are

$$T^{'} = \frac{T^0 - T_m^{\ 0}}{T_h^{\ 0} - T_c^{\ 0}}, S^* = \frac{S^0}{c_1(T_h^{\ 0} - T_c^{\ 0})}, C^* = \frac{C^0}{c}, K^* = \frac{k}{k_1}$$

$$Ste = \frac{c_1(T_h^{\ 0} - T_c^{\ 0})}{L}, C_{sl} = \frac{c_s}{c_1}, K_{sl} = \frac{k_s}{k_l} \quad (8)$$

It is important to know the characteristics of the laser beam profile. Generally, the spatial distribution of the laser beam intensity is represented as either a top-hat profile or Gaussian, while the temporal dependence may often be approximated as constant or Gaussian profile. The general laser beam intensity is represented as

$$I(r,t) = I_0 \exp[-\frac{(t-t_n)^2}{t_0^2}]\exp[-\frac{(r-r_c)^2}{r_0^2}] \quad (9)$$

where $I_0$ is the peak value of beam intensity.

By integrating the intensity over the beam area in space and the pulse duration in time, one can obtain the total amount of energy delivered by the laser beam as follows:

$$E = 2\pi \int_0^t dt \int_0^R rdr I(r,t) \quad (10)$$

As the beam penetrates into the target metal material, the electromagnetic energy is absorbed and resulting in damping of the intensity occurs over a very shallow depth of the material. The energy deposition is considered by assuming that all the energy is deposited into the top surface of the target material as a source on the surface. It is assumed that the surface temperature of the target material is high enough so that the reflectivity can be neglected.

Temperature is considered to be continuous across the melt/vapor region. The melt surface properties are determined from the conservation of mass, momentum and energy fluxes across the melt/vapor interface. The mass, momentum and energy balance across the melt/vapor interface with respect to a moving frame can be written as follows [17]

$$\rho_m u_m = \rho_v u_v \quad (11)$$

$$p_m + \rho_m u_m^2 = p_v + \rho_v u_v^2 \quad (12)$$

$$I_{abs} - L_v \rho_v u_v - k\frac{\partial T}{\partial n}\bigg|_{s/melt} = 0 \quad (13)$$

where $I_{abs}$ is the rate of energy absorption per unit surface area. Some previous studies indicated that the gas velocity leaving from the surface is considered as sonic at the laser intensities typical of laser drilling. The surface pressure and temperature are related by Clausius-Clapeyron equation and ideal gas law:

$$p(T_S) = p_{vap,0} \exp[\frac{L_v}{R^{'}}(\frac{1}{T_{vap,0}} - \frac{1}{T_S})] \quad (14)$$

$$p_v = R^{'} \rho_v T_v \quad (15)$$

Applying ideal gas law in the combined equation of the energy equation (13), Clausius-Clapeyron equation (14), we get

$$\frac{\gamma+1}{L_v \gamma}\sqrt{\gamma R T_s}(I_{abs} - k\frac{\partial T}{\partial n}\bigg|_{s/melt}) = p_{vap,0}\exp[\frac{L_v}{R^{'}}(\frac{1}{T_{vap,0}} - \frac{1}{T_S})] \quad (16)$$

Temperature gradient may be avoid for the high beam intensities due to the less conduction in melt region where Eq. (16) can be approximate with $\frac{\partial T}{\partial n}\bigg|_{s/melt} \approx \frac{I_{abs}}{k}$.

It is necessary to have the boundary conditions at mesh boundaries and at the free surfaces. Layer of artificial cells is enforced to the different boundary conditions. Zero normal component of the velocity and zero normal gradients tangential velocity are considered. The left boundary is assumed to be a no-slip rigid wall which results zero tangential velocity component at the wall. Normal stressed boundary condition is applied to the free surface. The surface cell pressure is calculated by a linear interpolation between impressed pressure on the surface and the pressure inside the fluid of the adjacent full fluid cell. Adiabatic boundary conditions are applied at the left, right and bottom boundaries. Stefan boundary condition is applied to solve the problem where the temperature of melt/vapor is unknown a *priori*. The target material is considered as ambient temperature at the beginning where the top surface of the substrate is considered as free surface liquid cells.

## 3. NUMERICAL SIMULATIONS

Volume of fluid (VOF) method is used to obtain the location of the free surface. The resultant melt velocity field is used to solve the energy equation to obtain the temperature field at the same time step. The velocity and pressure filed is solved for the free surface. The temperature field is solved by using control volume finite difference method [21] for the phase front as well.

### 3.1 Velocity and pressure calculation

VOF is a free surface modeling numerical technique and it is used for tracking the free surface. It refers to the Eulerian methods which are characterized by a mesh that is either stationary or moving in a certain manner to accommodate the shape of the interface. A rough shape of free surface is produced from the upstream and downstream values of *F* of the flux boundary. This shape is then used to calculate the boundary flux using pressure and velocity as primary dependent variables. If the cell has nonzero value of *F* and at least one neighboring empty cell is defined as a free surface cell. The velocities for (n+1)[th] interval is calculated from the pressure occurring at the same (n+1)$^{th}$ interval. The pressure iteration from the continuity equation is carried out until it is satisfied the implicit relationship between pressure and velocity for all the fluid cells. Pressure iteration should also satisfy the surface pressure boundary condition for all the free surface. The conservation of mass is maintained when applying free boundary condition for the free surface cells. The pressure which is the product of the surface tension coefficient and local curvature in each boundary cells is imposed on all the interfaces.

**3.2 Temperature field**

Finite volume method (FVM) is used to discretize the nonlinear energy equation. The iterative procedure requires for the solution as the energy equation is nonlinear due to the incorporation of phase change capability. At the beginning, the velocity for fixed grid at each time step is obtained. Those velocities are used in the advection terms of the energy equation to obtain the temperature field for the same fixed grid. The location of the temperature field is at the center of the cell where velocity components are located at the middle of the grid points on the control volume in the staggered grids.

VOF method is basically a finite difference method but to handle the donor-acceptor cell approximation a special function of F which results free surface location is used. So to handle both methods a combined single expression with a variable parameter which controls the relative amount of each is applied in the problem. It is shown that the location of the velocity variables in the control volume is same as VOF method because the VOF method was developed precedes the development of the control volume finite difference method.

As pressure and temperature go together, the free surface temperature boundary condition resulting from the gas dynamics and the pressure boundary condition are applied in the free surface. The velocity at the solid-liquid interface is attained by defining the kinematic viscosity as a function of temperature. The value of kinematic viscosity at liquid region is defined as the value of fluid viscosity and then gradually increased through the mushy zone to a large value for the solid region. No slip velocity boundary conditions are applied implicitly at the solid-liquid interface which can be easily implemented in the solution algorithm.

**4. RESULTS AND DISCUSSIONS**

The LD model treats the coupled problem consists of convection and conduction heat transfer; phase change processes (melting, solidification and vaporization), time- and position-dependent temperature and pressure which develop at the melt/vapor interface, and incompressible laminar flow of the melt with a free surface. The computer code has been used to solve two dimensional axisymmetric LD simulation using Hastelloy-X as a target material. Laser drilling on a Hastelloy-X workpiece is simulated and results are compared (Fig. 2) with the experimental data and calculated data from 2-D model in [18].

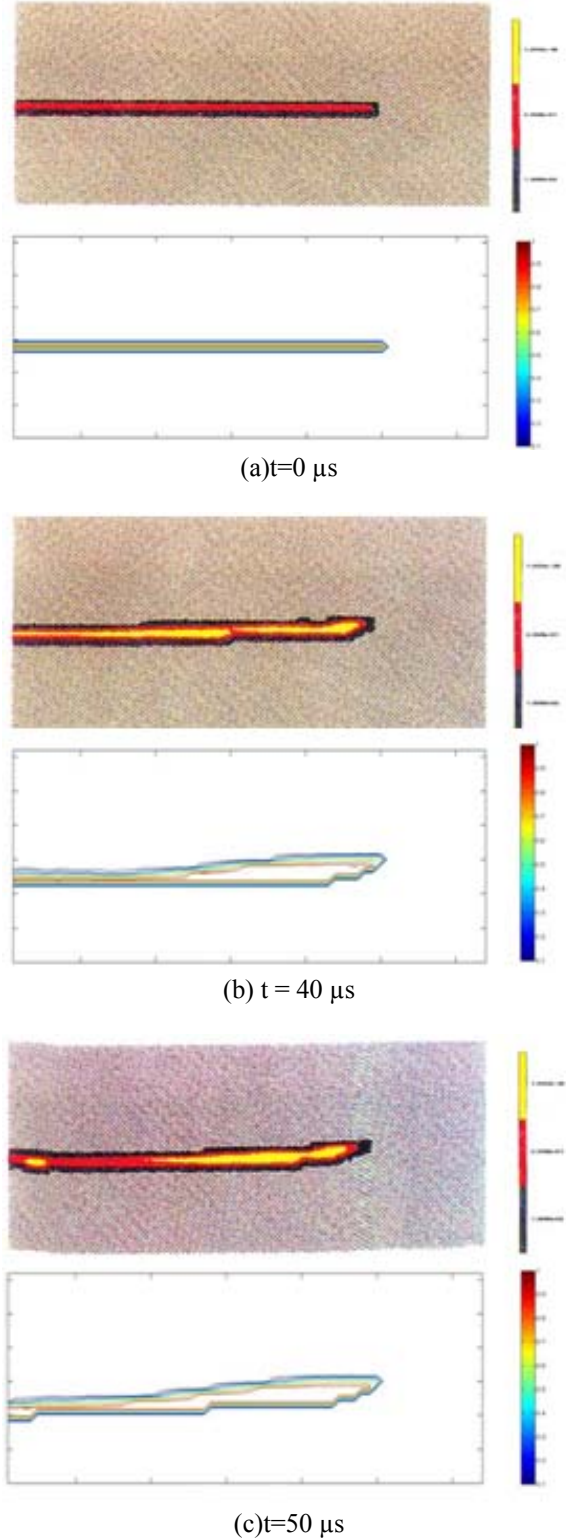

(a) t=0 μs

(b) t = 40 μs

(c) t=50 μs

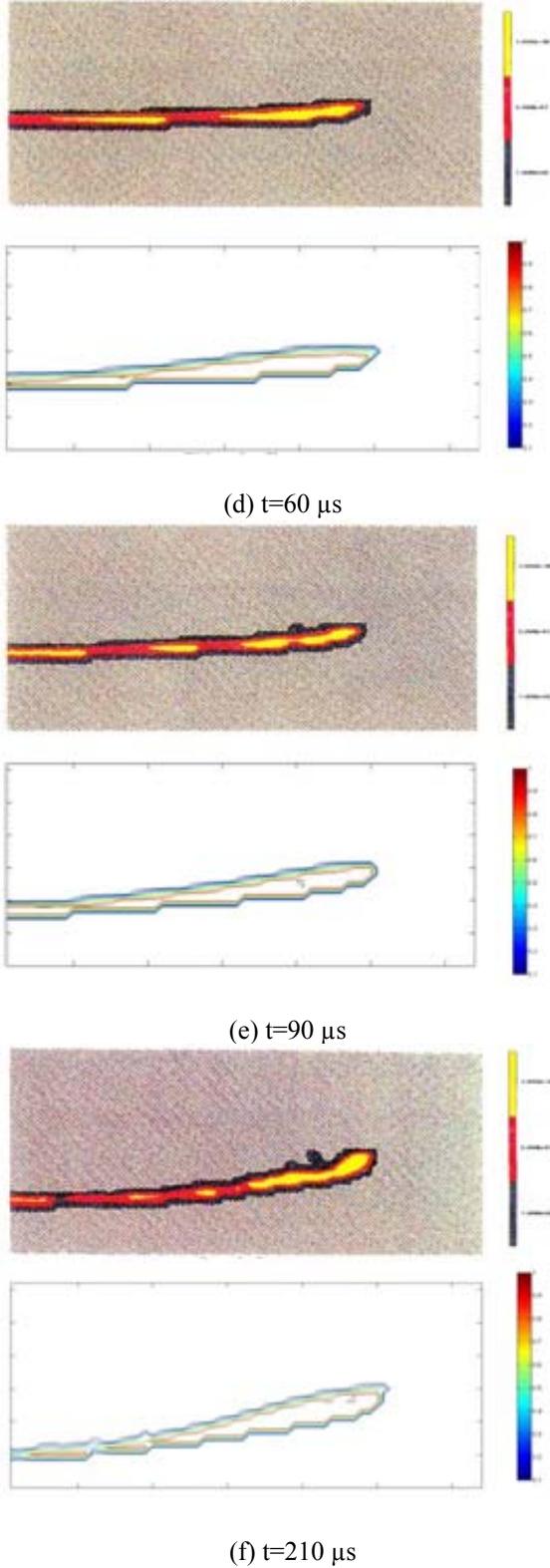

(d) t=60 μs

(e) t=90 μs

(f) t=210 μs

**Figure 2** Comparison of the fluid contour of the literature (top) and the current result (bottom) at different time sequence

The thermophysical properties of Hastelloy-X are given in Table 1.

**Table 1:** Thermophysical properties of the Hastelloy-X

| Property | Symbol | Value |
|---|---|---|
| Thermal conductivity of melt | k | 21.7 W/m.k |
| Density of melt | $\rho_m$ | $8.4 \times 10^3$ kg/m$^3$ |
| Vaporization Pressure, | $p_{vap,0}$ | $1.013 \times 10^5$ pa |
| Specific heat of melt | $c_p/c$ | 625 J/Kg.K |
| Temp. of vaporization | $T_{v,0}$ | 3100 K |
| Temp. of melt | $T_m$ | 1510 K |
| Latent heat of vaporization | $L_v$ | $6.44 \times 10^6$ J/g |
| Latent heat of melt | $L_m$ | $2.31 \times 10^5$ J/g |
| Molar mass | M | 76 g/mol |
| Dynamic viscosity | η | 0.05 g/cm.s |
| Surface tension | $\gamma_{ST}$ | 0.0001 J/cm$^2$ |
| Prandtl number | Pr | 0.142 |
| Schmidt number | Sc | 0.27 |
| Gas constant | R | 109 J/kg.K |
| Thermal diffusivity of melt | ƙ | $4.2 \times 10^{-6}$ m$^2$/s |

The diameter of the laser beam is 508 $\mu m$, which is also the length of the solid in the radial direction (100 cells). In addition, there are 25 cells of solid and 25 cells of air (empty) in the axial direction. Therefore, the length of the contour (in the radial direction) is 254 $\mu m$ with 50 cells. There are 25 empty cells located on the top of the solid cells. Each cell represents as 5.08 µ$m$ by 5.08µm square. Figure 3 shows the fluid contour at different times where the fluid cells are marked by values ranging from 0 to 1. The sequence of fluid contour illustrates the radial movement of the melt caused by the pressure gradient and its ejection.

### 4.1 Effects of laser beam diameter

Starting from the case discussed in Fig. 3, we considered several additional cases by changing the beam diameter from 508μm with the same $I_{max}$ and some cases with the same beam diameter but changing the laser intensity for study the effect of beam size, laser intensity. Figure 3 shows fluid contour at different time sequence for the original case (d=508 µm and $I_{max}$= 1 MW/cm$^2$). Figure 4 shows the temperature contour for the original case.

Figures 5 and 6 represent the fluid and temperature contours for the case with laser diameter of 1.5 mm (3 times to the original diameter) and the same maximum laser intensity ($I_{max}$= 1 MW/cm$^2$). It is shown from the Figs. 3 and 5 that the ablation effect decreases with the increase of the laser diameter under the constant laser intensity and laser pulse.

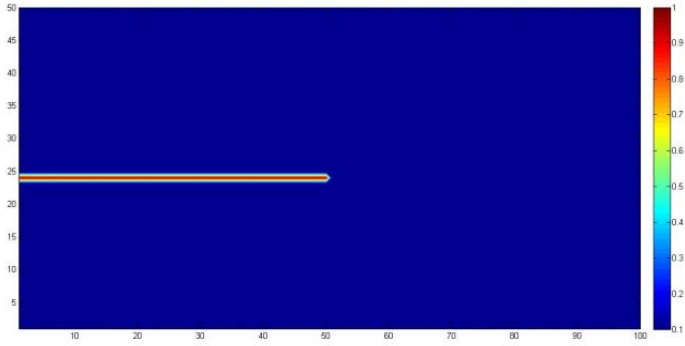

(a) t=0μs

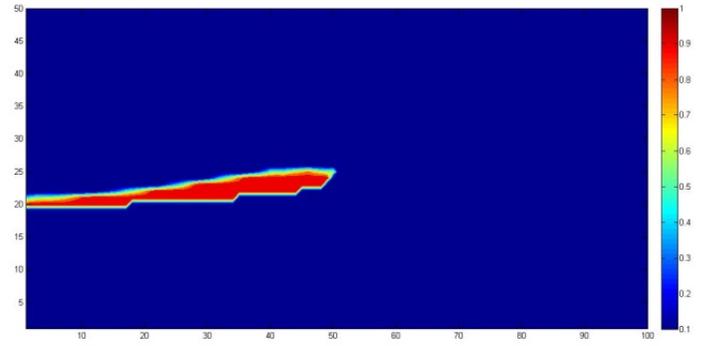

(d) t=60μs

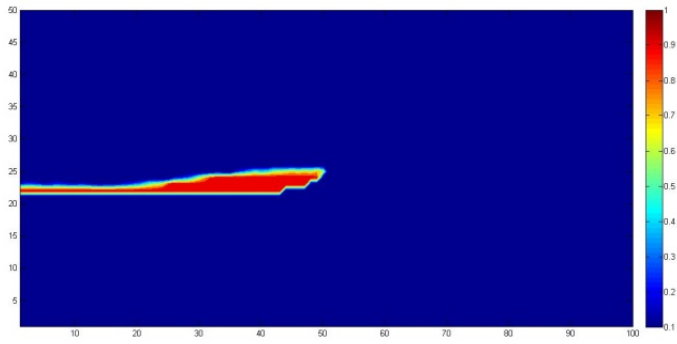

(b) t=40μs

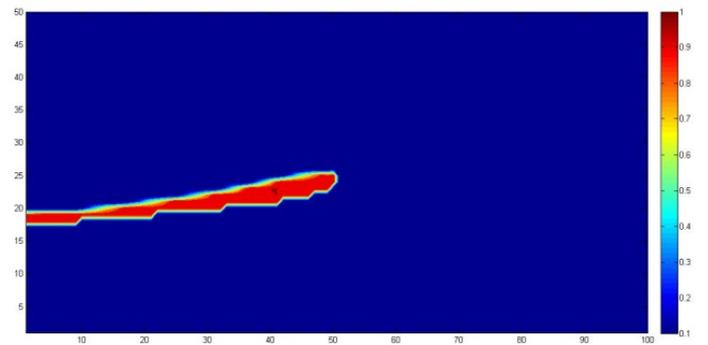

(e) t=90μs

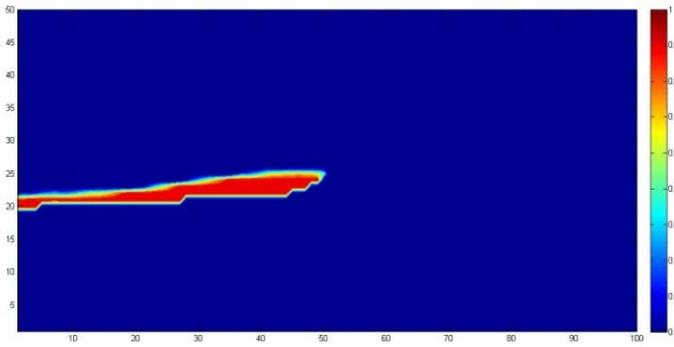

(c) t=50μs

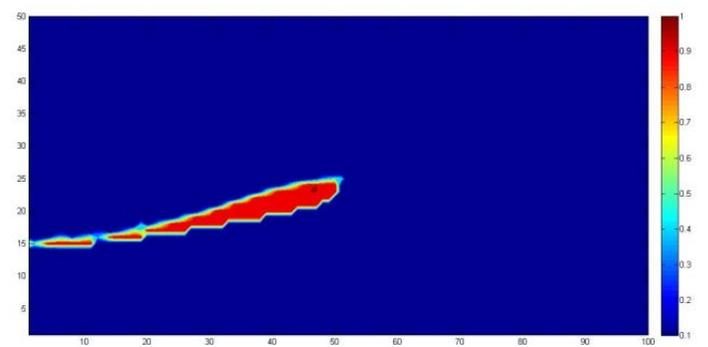

(f) t=210μs

**Figure 3** Fluid contour with $t_p = 210 \ \mu s$, R= 508μm and $I_{max} = 1 \ MW/cm^2$

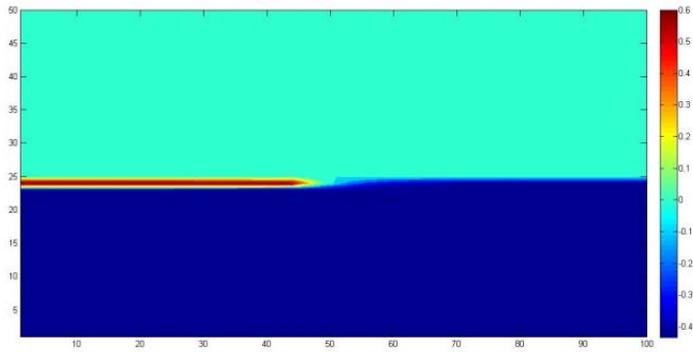

(a) t=0μs

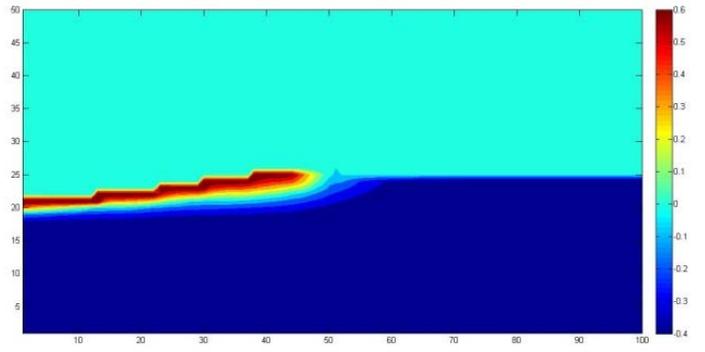

(d) t=60μs

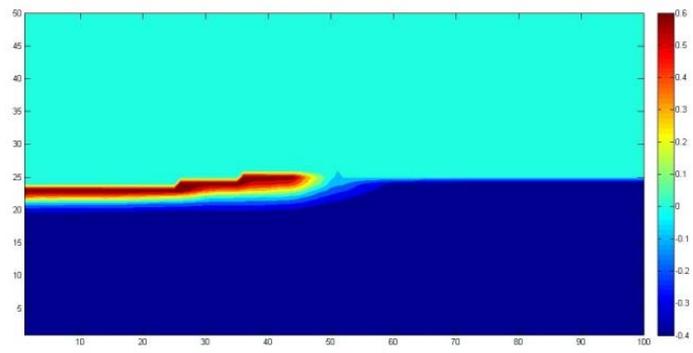

(b) t=40μs

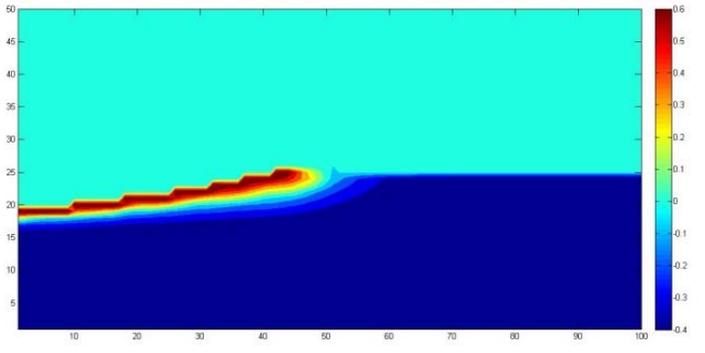

(e) t=90μs

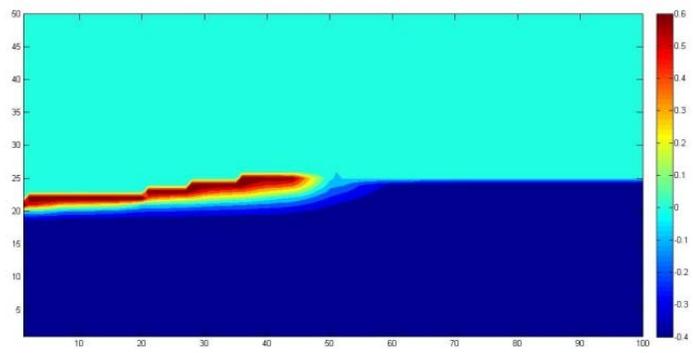

(c) t=50μs

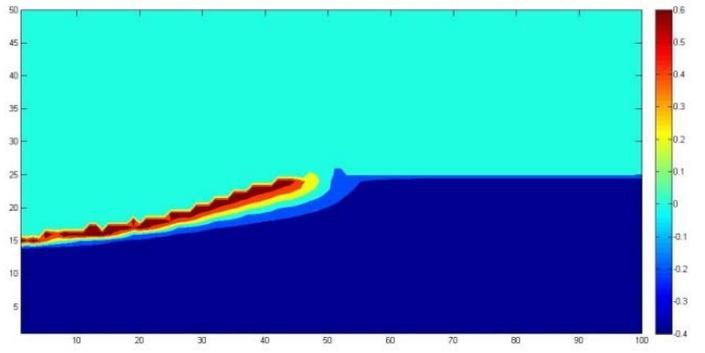

(f) t=210μs

**Figure 4** Temperature contours with $t_p = 210\ \mu s$, R= 508μm and $I_{max} = 1\ MW/cm^2$

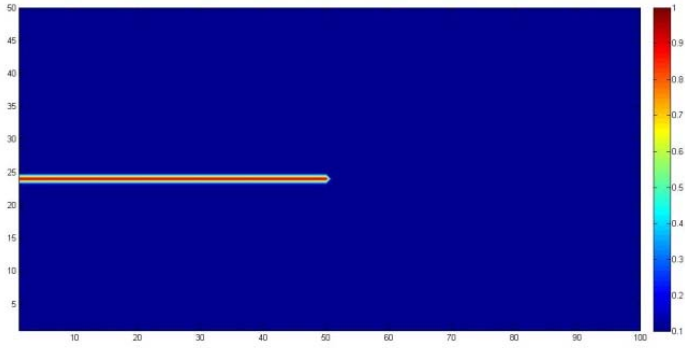

(a) t=0μs

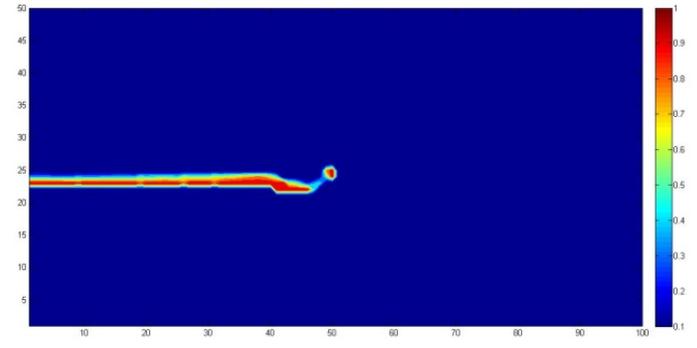

(d) t=60μs

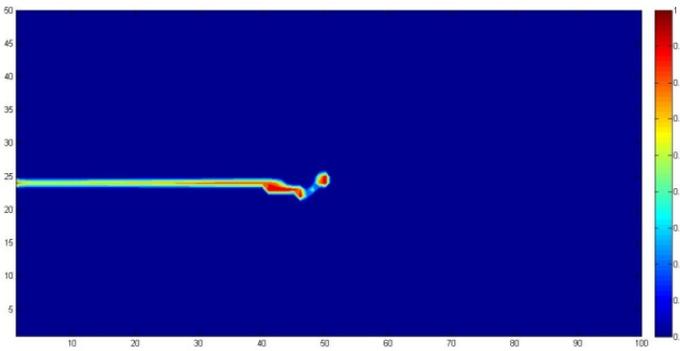

(b) t=40μs

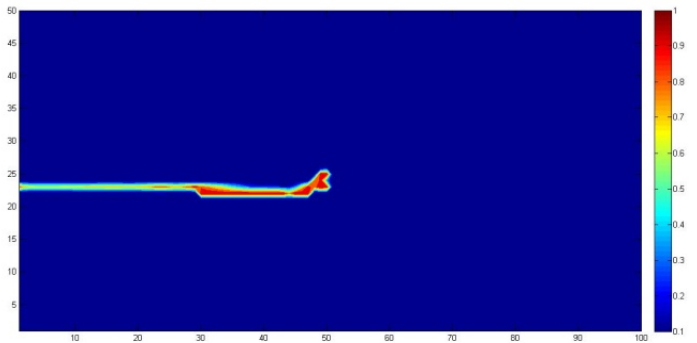

(e) t=90μs

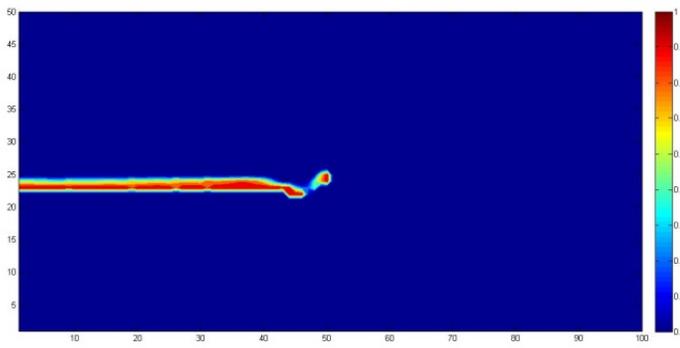

(c) t=50μs

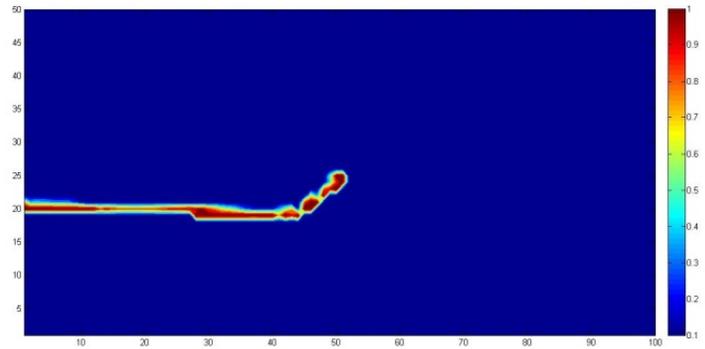

(f) t=210μs

**Figure 5** Fluid contour with $t_p = 210\ \mu s$, R= 1.5 mm and $I_{max} =\ 1\ MW/cm^2$

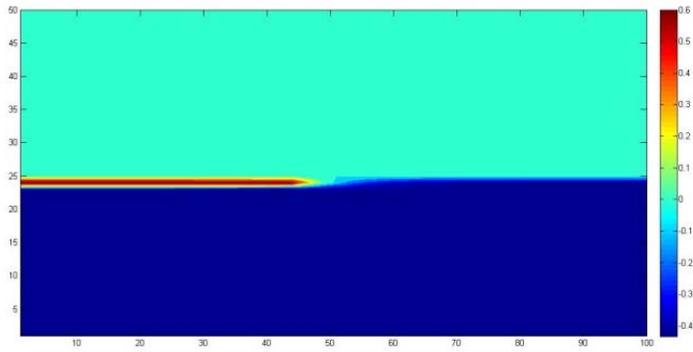

(a) t=0μs

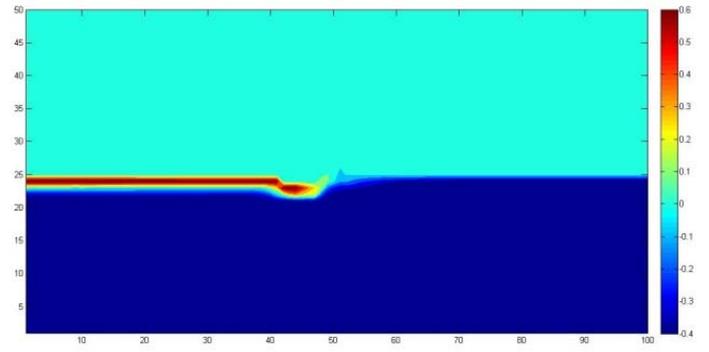

(d) t=60μs

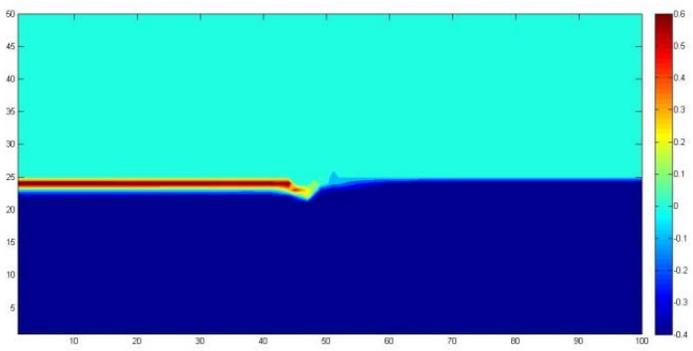

(b) t=40μs

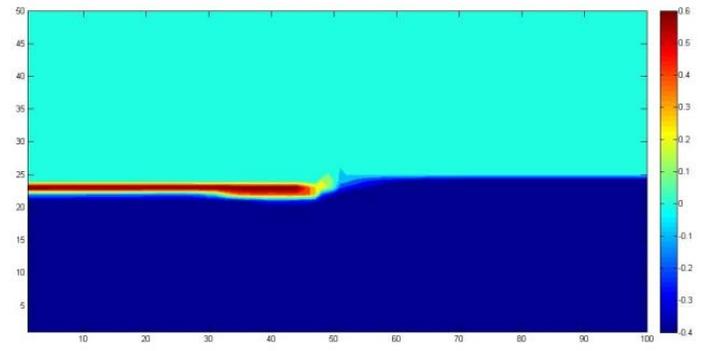

(e) t=90μs

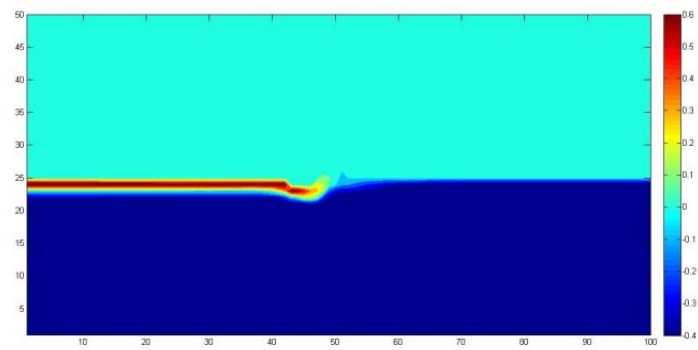

(c) t=50μs

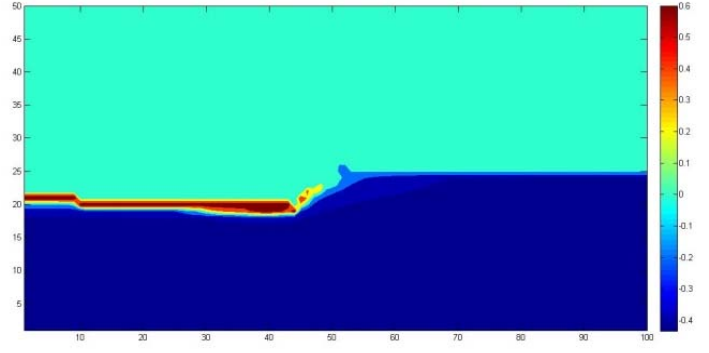

(f) t=210μs

**Figure 6** Temperature contours with $t_p = 210\ \mu s$, R= 1.5 mm and $I_{max} = 1\ MW/cm^2$

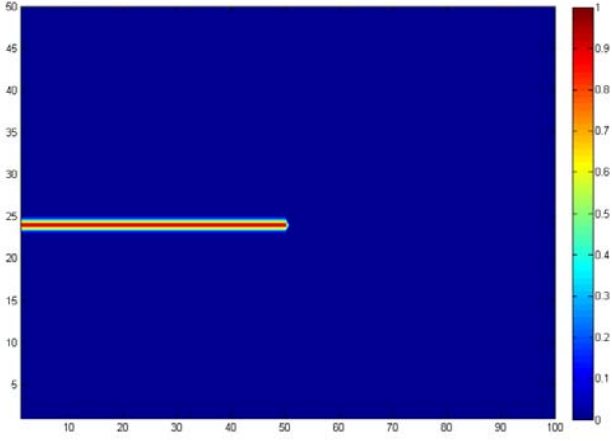

(a) $t = 0\ \mu s$

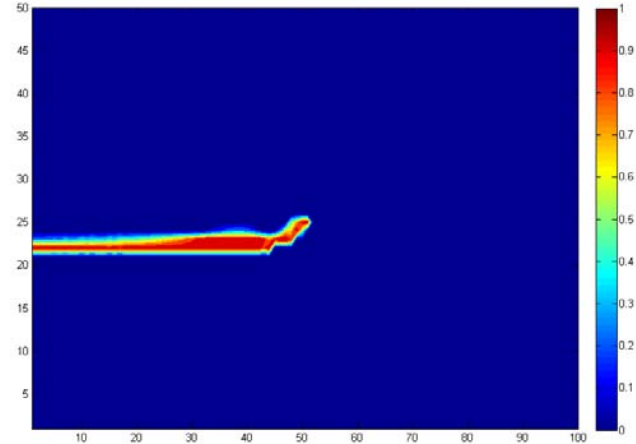

(c) $t = 25\ \mu s$

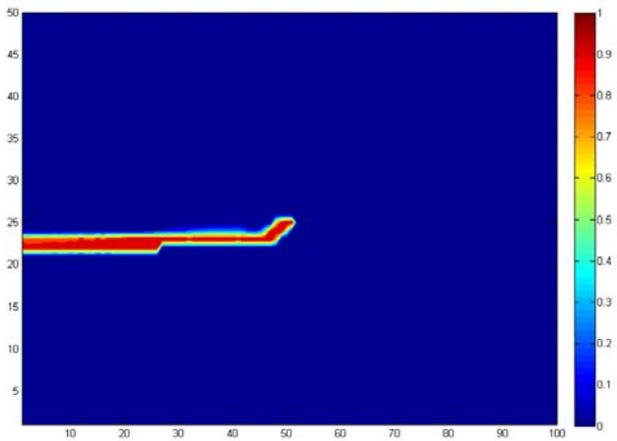

(b) $t = 20\ \mu s$

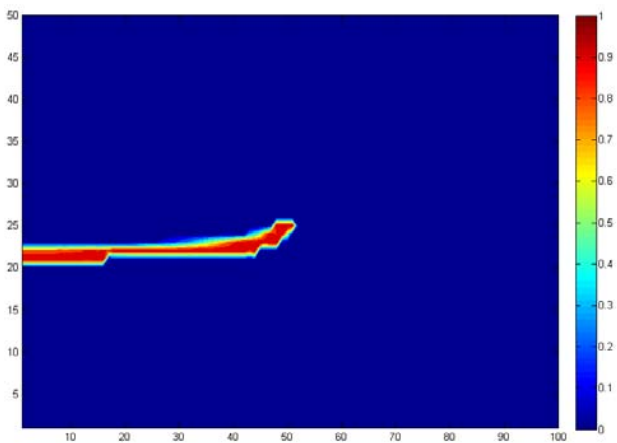

(d) $t = 30\ \mu s$

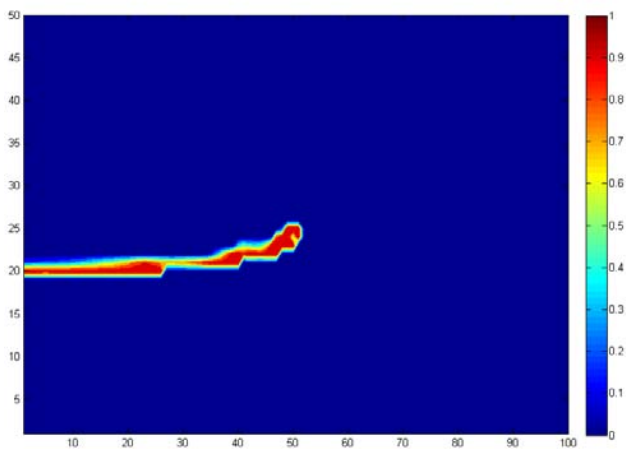

(e) $t = 45\ \mu s$

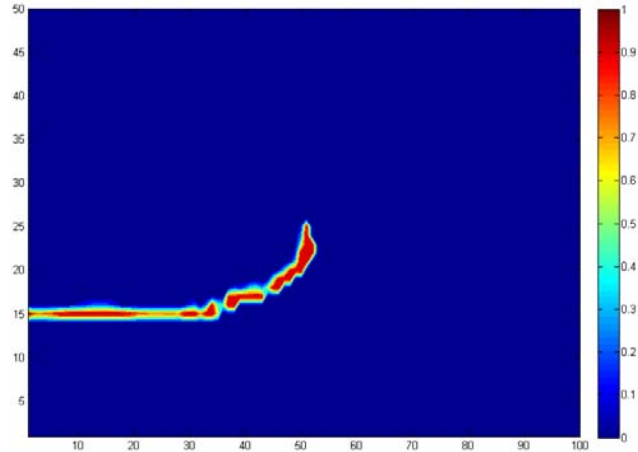

(f) $t = 105\ \mu s$

**Figure 7** Fluid contour with $t_p = 105\ \mu s$, $I_{max} = 2\ MW/cm^2$ and $D = 508\ \mu m$

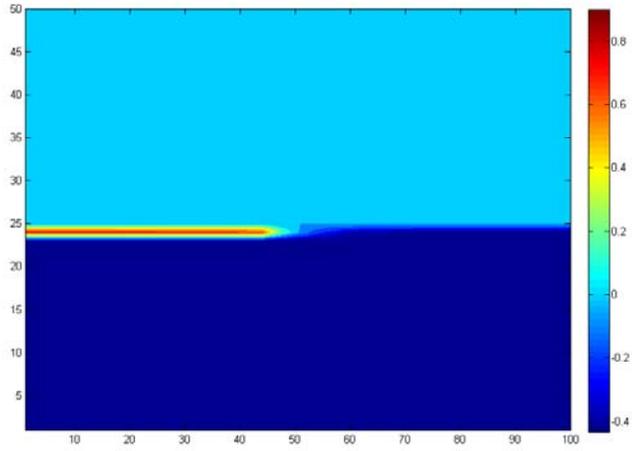

(a) $t = 0\ \mu s$

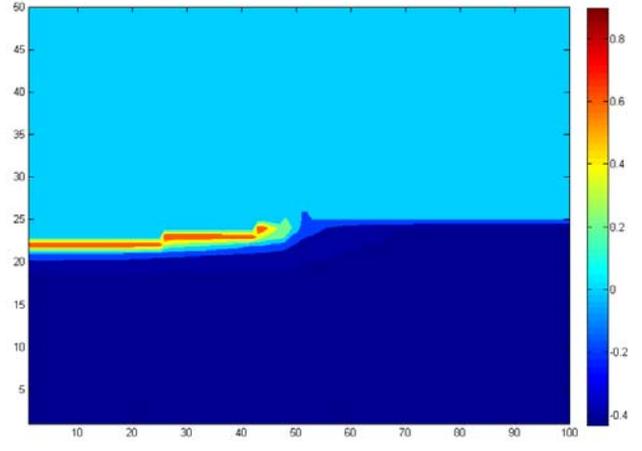

(d) $t = 30\ \mu s$

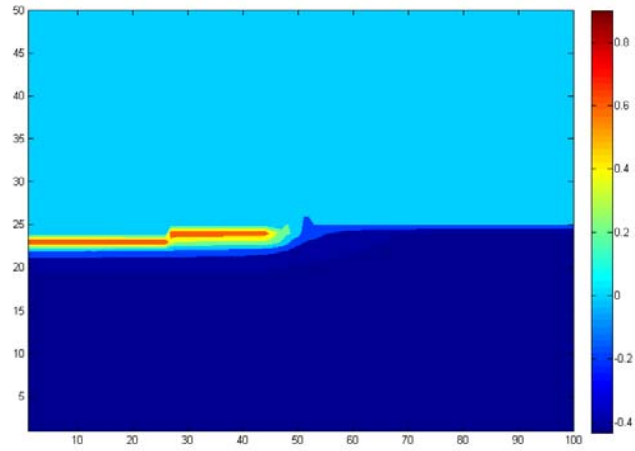

(b) $t = 20\ \mu s$

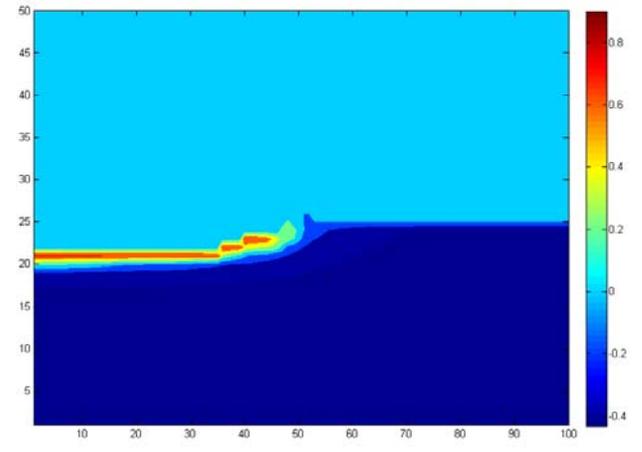

(e) $t = 45\ \mu s$

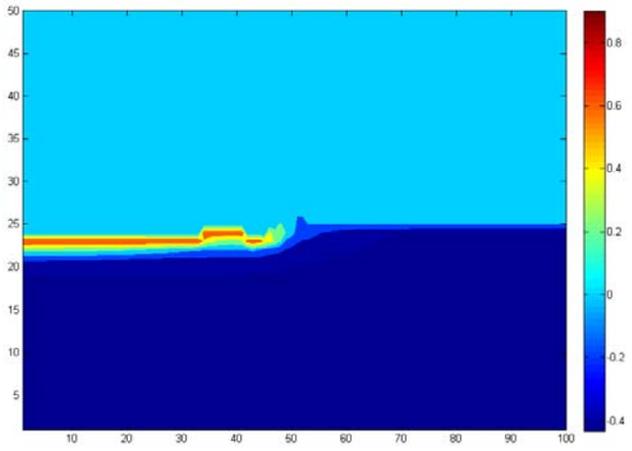

(c) $t = 25\ \mu s$

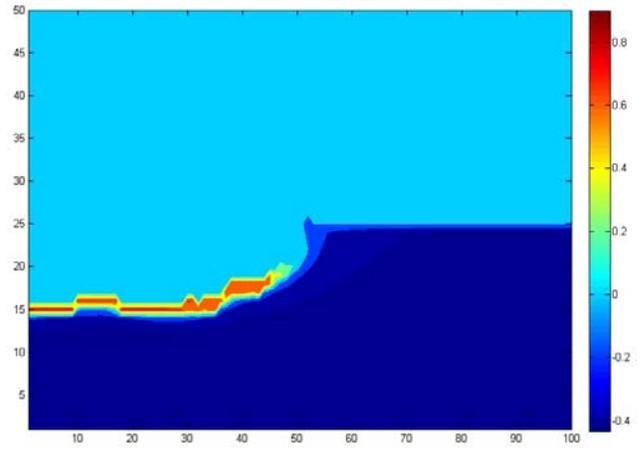

(f) $t = 105\ \mu s$

**Figure 8** Temperature contours with $t_p = 105\ \mu s$, $I_{max} = 2\ MW/cm^2$ and $D = 508\ \mu m$

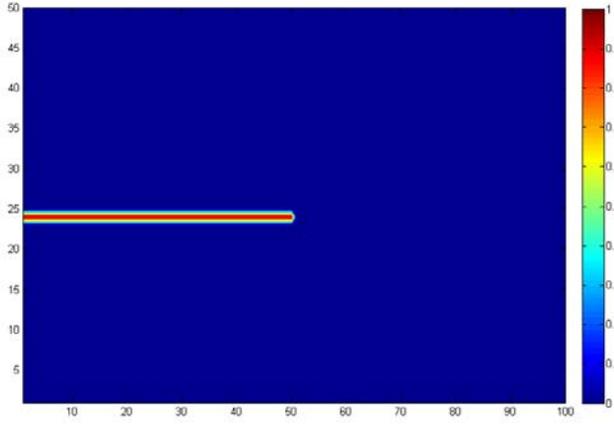
(a) $t = 0\ \mu s$

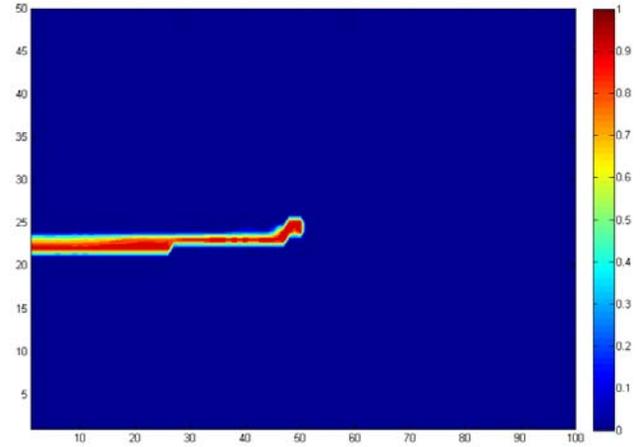
(d) $t = 15\ \mu s$

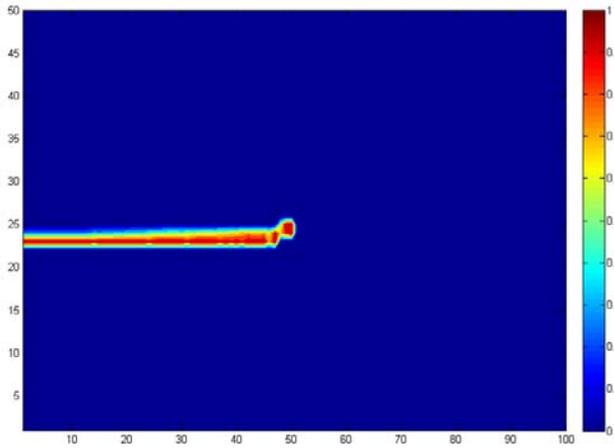
(b) $t = 10\ \mu s$

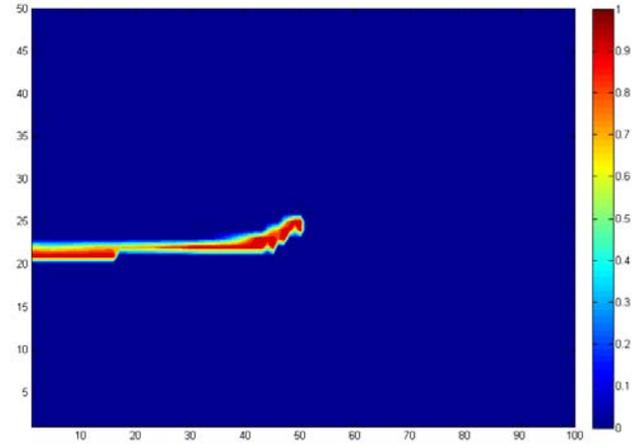
(e) $t = 22.5\ \mu s$

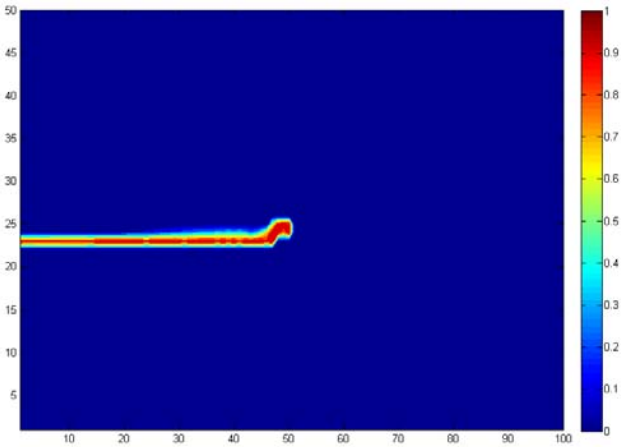
(c) $t = 12.5\ \mu s$

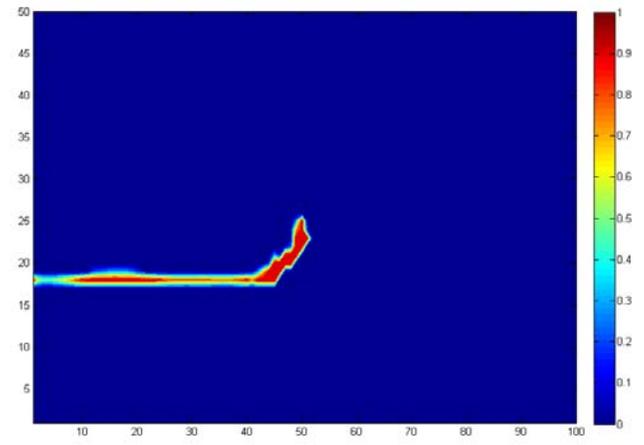
(f) $t = 52.5\ \mu s$

**Figure 9** Fluid contour with $t_p = 52.5\ \mu s$, $I_{max} = 4\ MW/cm^2$ and $D = 508\ \mu m$

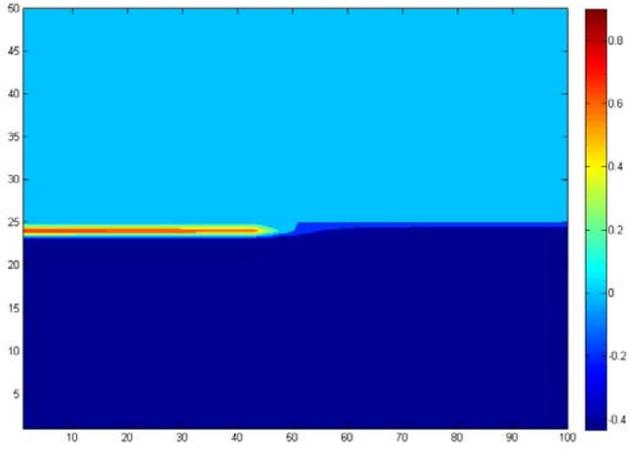

(a) $t = 0\ \mu s$

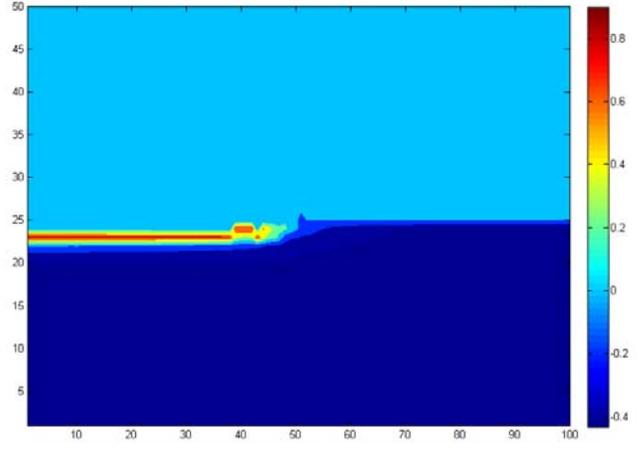

(d) $t = 15\ \mu s$

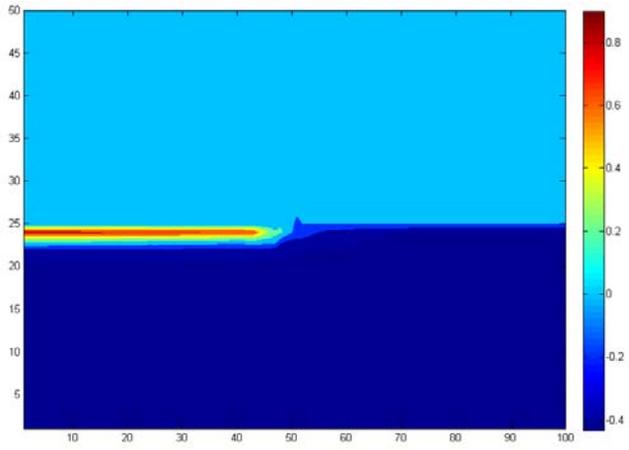

(b) $t = 10\ \mu s$

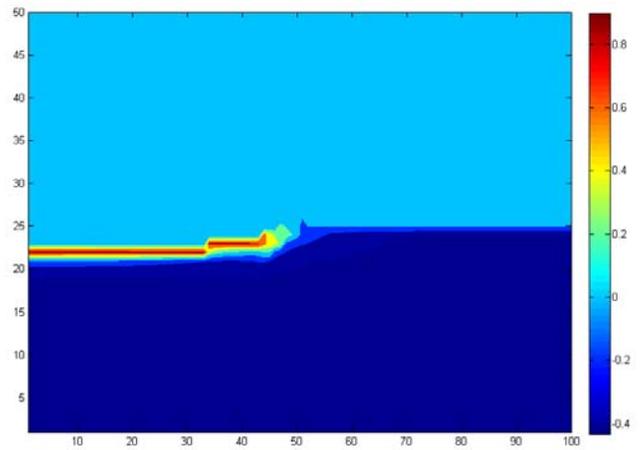

(e) $t = 22.5\ \mu s$

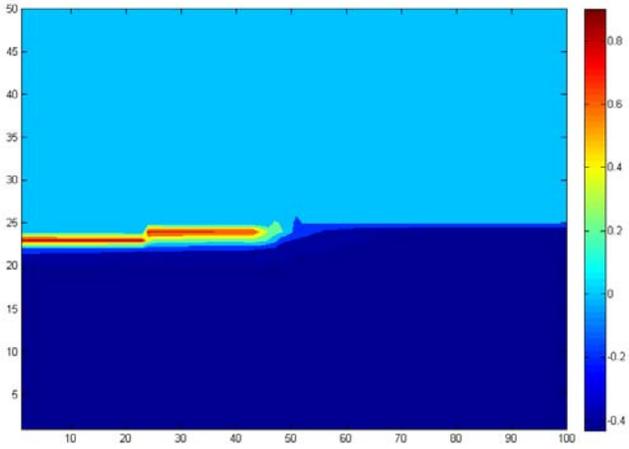

(c) $t = 12.5\ \mu s$

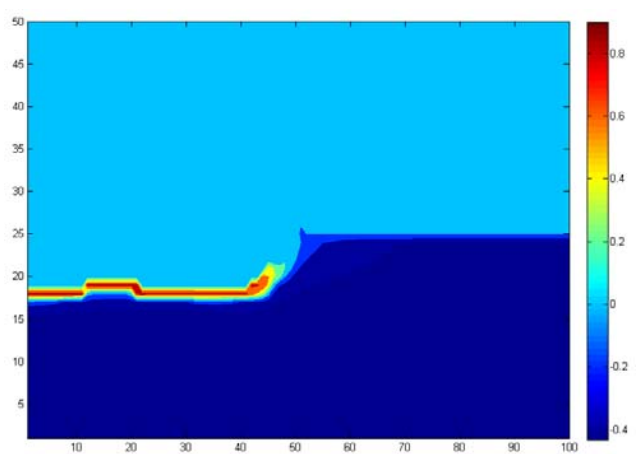

(f) $t = 52.5\ \mu s$

**Figure 10** Temperature contours with $t_p = 52.5\ \mu s$, $I_{max} = 4 MW/cm^2$ and $D = 508\ \mu m$

Although a deeper hole should observed due to the higher the laser power for the beam diameter D=1.5mm than the of the beam diameter D=508 µm, we found a shallow depth for the increased beam diameter. The reason behind is that under the same laser pulse width and the laser intensity, the increase of beam diameter results increased vaporization rate and then a thin layer of molten layer appeared. Another reason should be the validity of the application of Clausius/Clapeyron equation in this model. Under high pressure and near the critical point, Clausius/Clapeyron equation will give inaccurate results. Another reason may occur due to the application of Eq. (16). During the derivation of Eq. (16), we assumed the pressure just below the melting region is same as surface pressure and the surface temperature is same as the melting temperature at the melt-solid region.

### 4.2 Effects of laser pulse duration

The effects of laser pulse are then investigated. Fluid and temperature contours are shown in Figs.7 and 8 for the pulse duration of 105µs and maximum intensity of 2MW/cm$^2$ with original beam diameter 508µm. It is seen from the figures that the penetration decreases as the pulse duration decreases. Figures 9 and 10 represent the fluid and temperature contours for the case with pulse duration of 52.5µs and maximum intensity of 4MW/cm$^2$. Comparing the fluid contour plots in Figs. 7 and 9, it is shown that the hole diameter decreases with the decrease of laser pulse.

### 5. CONCLUSION

Effects of laser pulse width and beam diameter on laser drilling is investigated. The cases where the laser diameter changed from the original case (d=508µm) with the same maximum laser intensity are studied. It is shown that the hole depth increase with the decrease of beam diameter. The pulse duration effects with different laser intensities are also studied here. The pulse duration study concludes that when the laser pulse duration increases, the depth of the hole increases. Those discriminations should be introduced due to the application of Clausius/Clapeyron equation and some assumptions that had been taken in the presentation of pressure and temperature at the melt-solid region.

### NOMENCLATURE

| | |
|---|---|
| A | Nondimensional thermal diffusivity |
| B | Source term in the nonlinear energy equation |
| c | Specific heat (J/Kg.K) |
| $C^0$ | Coefficient of the nonlinear term |
| $d_0$ | Characteristic length |
| E | Enthalpy (J/kg) |
| F | Volume of fluid function |
| G | Nondimensional gravitational acceleration |
| g | Gravitational acceleration (9.8 m/s$^2$) |
| I | Intensity of the laser beam (W/m$^2$) |
| k | Thermal conductivity of melt(W/m.k) |
| $L_v$ | Latent heat of vaporization (J/g) |
| $L_m$ | Latent heat of melting (J/g) |
| M | Molar mass (g/mol) |
| P | Nondimensional pressure |
| $p_{vap,0}$ | Vaporization pressure (Pa) |
| Pr | Prandtl number |
| r | radial coordinate |
| R | Nondimensional radius ( m) |
| R' | Gas constant (J/kg.K) |
| $S^0$ | Coefficient |
| Sc | Schmidt number |
| t | Time coordinate |
| T | Nondimensional time |
| $T^0$ | Temperature |
| $T_{v,0}$ | Temperature of vaporization (K) |
| $T_m^0$ | Temperature of melting (K) |
| U | Nondimensional velocity in x-direction (m/s) |
| u | Velocity in x-direction(m/s) |
| V | Nondimensional velocity in y-direction (m/s) |
| v | Velocity in y-direction (m/s) |
| Z | Nondimensional axial coordinate |
| z | Axial coordinate |

Greek Symbols

| | |
|---|---|
| α | Thermal diffusivity (m$^2$/s) |
| η | Dynamic viscosity (g/cm.s) |
| σ | Surface tension coefficient (N/m) |
| $γ_{ST}$ | Surface tension (J/cm$^2$) |
| ǩ | Thermal diffusivity of melt (m$^2$/s) |
| ρ | Density (kg/m$^3$) |

* represents the dimensionless